# Lonsdaleite: The diamond with optimized bond lengths and enhanced hardness

Liuxiang Yang[1], Kah Chun Lau[2], Zhidan Zeng[1], Dongzhou Zhang[3], Hu Tang[1], Bingmin Yan[1], Huiyang Gou[1], Yanping Yang[1], Yuming Xiao[4], Duan Luo[5], Srilok Srinivasan[5], Subramanian Sankaranarayanan[5,6], Wenge Yang[1,*], Jianguo Wen[5,*], Ho-kwang Mao[1,*]

**Affiliations:**

[1]Center for High Pressure Science and Technology Advanced Research, Beijing 100193, P.R. China

[2]Department of Physics and Astronomy, California State University Northridge, Northridge, California 91330, USA

[3]Hawai'i Institute of Geophysics and Planetology, School of Ocean and Earth Science and Technology, University of Hawai'i at Manoa, Honolulu, Hawaii 96822, USA

[4]HPCAT, Geophysical Laboratory, Carnegie Institution of Washington, Argonne, Illinois 60439, USA

[5]Center for Nanoscale Materials, Argonne National Laboratory, Argonne, Illinois 60439, USA

[6]Department of Mechanical and Industrial Engineering, University of Illinois, Chicago, IL 60607 USA

*Corresponding authors: yangwg@hpstar.ac.cn (W.Y.), jwen@anl.gov (J. W.) and maohk@hpstar.ac.cn (H.K.M.)

**Diamond is known as the hardest substance due to its ultra-strong tetrahedral $sp^3$ carbon bonding framework. The only weak link is its cubic cleavage planes between (111) buckled honeycomb layers. Compressing graphite single crystals and heating to moderate temperatures, we synthesized a bulk, pure, hexagonal diamond (lonsdaleite) with distorted carbon tetrahedrons that shorten the bond between its hexagonal (001) buckled honeycomb layers, thus strengthening their linkage. We observed direct transformation of graphite (100) to lonsdaleite (002) and graphite (002) to lonsdaleite (100). We find the bulk lonsdaleite has superior mechanical properties of Vicker hardnesses $H_V$ = 164±11 GPa and 124±13 GPa, measured on the surface corresponding to the original graphite (001) and (100) surfaces, respectively. Properties of lonsdaleite as the supreme material can be further enhanced by purifying the starting material graphite carbon and fine-tuning the high pressure-temperature synthesis conditions.**

Diamond possesses unmatched hardness and other extreme properties, and is regarded as an ideal material for applications ranging from super abrasives, heat sink, bio-sensor, quantum computation, to photonic devices[1,2]. These supreme properties are originated from its unique building blocks of $sp^3$ carbon, bonded to each other with 1.54Å bond length and 109.5° bond angle that form perfect tetrahedrons. This single type of bonding extends infinitely in two dimensions to form buckled honeycomb carbon layers and the layers stack in the third dimension to form the (111) planes of cubic diamond crystals. Such structure, however, also has a weakness. The (111) linkage are relatively weak between layers, resulting in weak (111) cleavage planes that limit the strength of diamond. Selectively strengthening interlayer bonds relative to the intralayer bonds would lower the symmetry of cubic diamond to hexagonal.

Hexagonal diamond was predicted decades ago[3], and subsequently synthesized by dynamic explosion[4] and static compression[5]. Natural hexagonal diamond was discovered in Canyon Diablo iron meteorite, and named lonsdaleite after pioneer woman crystallographer Kathleen Lonsdale[6]. Although lonsdaleite was theoretically predicted to have superior mechanical properties than cubic diamond[7], pure, bulk lonsdaleite has never been synthesized and its hardness has thus never been measured.

Decades of relentless research efforts on lonsdaleite have led to dismal progresses in synthesis of lonsdaleite. It even remains unclear whether lonsdaleite actually is a discrete phase or a faulted, twined cubic diamond[8]. All previously known synthetic or natural lonsdaleite samples are fragile, fine-grain mixtures of hexagonal diamond with various proportions of cubic diamonds, amorphous carbon or residual graphite[6,7, 9-13]. Mechanical properties of lonsdaleite could not be delineated in such mixture of multiple phases.

Here we report comprehensive high pressure-temperature (*P-T*) studies to establish the existence of lonsdaleite as a bona fide phase, to determine its structure relation with its parental graphite, to synthesize pure bulk lonsdaleite without other phases, and to measure its mechanical properties. Both multi-anvil (MA) press and diamond-anvil cell (DAC) were used for compressing and heating single-crystal graphite starting material under quasi-hydrostatic high-*P* condition. Their crystallographic relationships were monitored *in-situ* at high *P* during the graphite-lonsdaleite conversion by X-ray diffraction (XRD) and optical observation. Pure lonsdaleite synthesized at high *P-T* were successfully recovered at ambient conditions for further *ex-situ* investigations using XRD, high resolution transmission electron microscopy (HRTEM), X-ray Raman spectroscopy (XRS), electron energy loss spectroscopy (EELS), UV Raman spectroscopy to and Vickers hardness measurements. We demonstrate that lonsdaleite is indeed the hexagonal counterpart of cubic diamond, with shortened and strengthened bonding between buckled honeycomb layers. These findings open new exploration of lonsdaleite as a potentially superior technological material.

We conducted 30 experiments in the past 6 years to investigate key issues of the graphite-lonsdaleite transition. We studied graphite single crystal sample (001 flak) in neon pressure-transmitting medium in a DAC to keep the samples intact before, during, and after the transition and to preserve the crystallographic orientations of the products and their relationship to the original graphite crystals. Fig. 1a (graphite at low pressure) shows a typical hexagonal [001] near zone-axis pattern with six (100) XRD spots equally distributed at 60° (azimuth angle) apart on a constant-2θ circle, and six (110) spots on the next larger constant-2θ circle with azimuth angles exactly in between of

(30° from) two adjacent (100) spots. The calculated *d*-spacings of (100) is exactly $\sqrt{3}$ times that of (110), reflecting the hexagonal honeycomb configuration of graphite. Above 20 GPa (Fig. 1b), the six-fold XRD pattern appears similar, except the (100) circle expands (*d*-spacing decreases) while the (110)-circle remains the same. This breaks the strict $\sqrt{3}$ rule and the graphite honeycomb must be broken and the six-fold G(100), G(010), G($\bar{1}$10), G($\bar{1}$00), G(0$\bar{1}$0), G(1$\bar{1}$0) spots convert to L(002), belonging to three distinctive sets of lonsdaleite nanocrystals, where G and L denote graphite and lonsdaleite, respectively. It was previously thought that the graphite [210] converted into lonsdaleite [001], and [010] was the common direction shared by the original graphite and resultant lonsdaleite[12,13]. With the advantage of following the single-crystal graphite honeycomb zone-axis pattern during the transition, we can now conclude that the correct relation is that the G[100] converts to L[001] and [110] is the common shared direction. We agree with the previously assignment of the third orientation relation of G[001] // L[100] [12,13].

The lonsdaleite sample, synthesized at 20 GPa and 1400 K, was retrieved at ambient conditions and the recovered sample was studied with XRD. The zone-axis pattern of the original graphite c-axis (Fig. 1c) in the recovered lonsdaleite specimen shows a six-fold XRD pattern similar to the XRD pattern at 20 GPa (Fig. 1b). By rotating the specimen, a series of diffraction spots were observed and the corresponding *d* values are listed in Table 1. All these peaks can be well indexed with a lonsdaleite phase (SG P6$_3$/mmc, *a*= 2.51 Å, *c*= 4.16 Å)[12,13]. The XRD patterns clearly show the crystallographic relation of single-crystal graphite and polycrystalline lonsdaleite as G[001] // L[100], G[100] // L[001] and G[110] // L[110] (Fig. 1.d,e). The flat carbon honeycomb layer (basal plane of the hexagonal crystal) of graphite does not convert into lonsdaleite honeycomb directly, but slides and links to the adjacent layers to form buckled carbon honeycomb layer of lonsdaleite in the perpendicular (100) directions of the original graphite (Fig. 1e). Due to the threefold degeneracy of the G{100} class, the original single-crystal graphite must split into a polycrystalline aggregate of three sets of lonsdaleite nanocrystals with L(002) corresponding to original G(100), G(0$\bar{1}$0), and G($\bar{1}$10).

To see the individual nanocrystals in the lonsdaleite specimen requires high-resolution transmission electron microscopy (HRTEM) at nm scale. Plane-view TEM specimen was prepared by precise cut in the original G(001) orientation from the recovered sample with focused ion-beam (FIB) technology. Selected-area electron diffraction (SAED) pattern (Fig. 1f) from the specimen using a 1 μm diameter aperture shows an ostensible 6-fold symmetry with *d* value of 2.08 Å, consistent with the XRD image in Fig. 1c. When the aperture diameter decreases to about 100 nm, the diffraction spots with only two-fold symmetry starts to appear (Fig. 1g) and it can be indexed by lonsdaleite [100] zone axis. Therefore, the hexagonal like XRD pattern (Fig. 1c) and the large SAED pattern (Fig. 1f) actually consist of three sets of lonsdaleite nanocrystals with c-axes parallel to the six symmetric directions of [100] in the original graphite single crystal. The SAED patterns and HRTEM images from several major zone axes of one lonsdaleite variant (in Supplementary Information Fig. SI-1) indicate that the recovered sample is a submicron aggregate of bona fide lonsdaleite phase with domain size around 100 nm.

The tetrahedral symmetry of the $sp^3$ bonded carbon in lonsdaleite is lowered to become a distorted trigonal pyramid geometry with: three OA bonds forming the buckled honeycomb layer, a OB bond linking two layers, three OA∧OA angles and three OA∧OB angles as shown by Fig. 2a. In order to study lonsdaleite structure on the actual bond length level, we proceeded to atomic resolution and utilized aberration-corrected HRTEM (AC-HRTEM) to image atom arrangement in a unit cell directly. Fig. 2b shows an AC-HRTEM image along the [110] zone axis and the carbon dumbbells with a projected separation of 0.89 Å is clearly distinguished. Both bonds OA and OB lay in the (110) plane and their bond lengths can be measured directly from the [110] zone axis HRTEM image. In the enlarged HRTEM image (Fig. 2c), we found that bond length OB (1.47 Å) is shorter than OA (1.56 Å). Based on the observation of two bond lengths and unit-cell parameters of lonsdaleite, we calculated OA∧OA=105.9° and OA∧OB=112.8°) (Fig. 2a), that is similar to the nano-domain structure reported by

Kanasaki et. al.[15].

Optical Raman spectroscopy provides valuable information on bonding length, angles, and strength that can often be used as finger prints for phase identification. Lonsdaleite has high luminescence background with optical visible excitations that often overwhelms the Raman signals. We overcame the challenge by using a 325 nm UV excitation laser[16] and obtained high quality lonsdaleite Raman spectra (Fig. 3a). Three peaks with vibration frequencies at 1321, 1529, and 1249 cm$^{-1}$ (in the order of peak intensity) have been observed. The absence of graphite G band at 1583 cm$^{-1}$ and D band at 1370 cm$^{-1}$, and the cubic diamond singular band at 1332 cm$^{-1}$ [17] indicates the absence of graphite and cubic diamond. Our peaks are in good agreement with previous measurements of lonsdaleite[16], but very different from theoretical simulation based on singular bond-length lonsdaleite structure that predicted three active Raman vibration modes at 1330.4, 1331.8 and 1218.4 cm$^{-1}$ [18,19]. The difference further substantiates that the carbon atoms arrangement in lonsdaleite deviates from the singular bond length model. Fig. 3b shows our DFT calculations (see Simulations in Methods) of lonsdaleite Raman vibration frequencies as a function of OB bond length with fixed OA at 1.56 Å. For OB at 1.44 Å, the three observed Raman peaks, 1321, 1529, and 1249 cm$^{-1}$ corresponding to $E_{1g}$, $A_{1g}$ and $E_{2g}$ Raman active modes, respectively, matching very well with the AC-HRTEM result in Fig. 3a. The $A_{1g}$ Raman active mode corresponds to the stretching mode of OB=1.44 Å, whereas $E_{1g}$ and $E_{2g}$ Raman active modes are attributed to stretching mode of OA=1.56 Å. Overall, optical Raman spectroscopic results are consistent with the OA/OB two bond length model in Fig. 2a and also indicate that the sample consist of only lonsdaleite.

Graphite-to-diamond transition is a change of carbon bonding from $sp^2$ $\pi$ bonds to $sp^3$ $\sigma$ bonds which can be diagnosed by electron energy-loss spectroscopy (EELS). We conducted EELS measurements during our TEM studies of lonsdaleite, and observed complete transition to $sp^3$ $\sigma$ bonds (Fig. 3c). EELS does not have penetration power and we use X-ray Raman spectroscopy (XRS), the energy loss spectroscopy using high-energy X-ray as the excitation source, to study bulk sample[20], and no sign of $\pi$ bond contribution had been detected (Fig. 3d), meaning all graphite $sp^2$ $\pi$ bonds have completely transformed to $sp^3\,\sigma$ bonds in the recovered lonsdaleite sample. EELS and XRS only diagnose $sp^2\,\pi$ and $sp^3\,\sigma$ bonds, but cannot differentiate lonsdaleite from cubic diamond, that have been well distinguish by XRD, HRTEM, and optical Raman spectroscopy.

With the demonstrated existence of bulk lonsdaleite sample, we proceeded to investigate the hardness of lonsdaleite, that required larger sample than DAC could produce. A 1 mm diameter lonsdaleite disk was synthesized from graphite crystal at 20 GPa and 2073 K in a large volume press. The details of sample preparation and hardness test are described in "Methods" section. The Vickers hardness measurements in two orientations are illustrated in the inset of Fig. 4. At 4.9 N load in both directions, the Vickers hardnesss-load curves of measurements parallel and perpendicular to c-axis of the original graphite reached asymptotic values of 164 ± 11 and 124±13 GPa, respectively. Both values are higher than that of natural diamond (~110 GPa) on the (110) surface under the same load[21]. The expected high hardness 164±11 GPa along the pristine graphite [001] direction which is perpendicular to the lonsdaleite [001], could be attributed to the shortened bond length (1.47 Å) and strengthened interlayer bonds along the *c*-axis of lonsdaleite (Fig. 2a), thus selectively eliminating the weakness of interlayer cleavage of cubic diamond.

In summary, our results unambiguously demonstrate the existence of lonsdaleite as a bona fide phase of carbon, and further experimentally confirm that lonsdaleite is a material with superior hardness possibly surpassing cubic diamond. Using an array of diagnostic techniques, we show that lonsdaleite has the optimal buckled honeycomb layered structure composing of distorted $sp^3$ tetrahedron units with regular 1.56 Å intralayer bond and shortened 1.47 Å interlayer bond that strengthens the linkage between layer relative to the cubic diamond structure. Very excitingly, this is just the beginning; lonsdaleite as a hexagonal variant of the cubic diamond has many extremely favorable mechanical, electrical, thermal and optical properties comparable and complementary to diamond. Lonsdaleite present great opportunities for exploration in the *P-T* conditions of synthesis, dopants in the precursor graphite, and nanocrystallinity to optimize the desirable specific features and to engineer the ideal *sp³* carbon allotrope.

## Methods

**Sample preparation and synchrotron characterization.** A high quality lonsdaleite has been synthesized under high pressure and temperature in a diamond anvil cell (DAC). The pristine material is a 60 μm (diameter) × 20 μm (height) disk cut from a millimeter size single crystal graphite by using the micro laser drilling system[23]. Then, it was loaded into a hundred micron diameter rhenium gasket chamber in a laser heating DAC. Pressure was monitored by ruby fluorescence method. When the applied pressure was above 20 GPa, and the sample was heated above 1400K with a YAG laser, the black graphite turned into transparent. Upon cooling and decompression to ambient conditions, this transparent sample can be recovered. We performed X-ray diffraction on the recovered sample at beamline16-BMD of High Pressure Collaborative Access Team (HPCAT) and 13-BMC of the GeoSoilEnviroConsortium for Advanced Radiation Sources (GSECARS), Advanced Photon Source (APS), Argonne National Laboratory (ANL).

**Preparation and Vickers Hardness measurements of large size lonsdaleite samples.** Large size samples (~1 mm in diameter, ~70 μm in thickness) were synthesized with a high-pressure multianvil apparatus by applying 20 GPa pressure and 2073 K temperature for 20 minutes. The COMPRES 8/3 assembly used in experiments consists of $MgAl_2O_4$ + MgO octahedron (pressure medium), a rhenium heater and a $LaCrO_3$ thermal insulator. In order to reduce the deformation of sample during compression, the graphite disk was surrounded by cubic boron nitride powder. Pressure was calibrated at room temperature based on the phase transition of pressure standard materials of GaAs (18.3 GPa) and GaP (23.0 GPa). Temperatures were monitored with a $W_{97}Re_3$-$W_{75}Re_{25}$ thermocouple. The recovered samples were checked by XRD to confirm the pure lonsdaleite phase. The hardness tests were conducted on the polished surfaces with a micro-Vickers hardness machine (FM-700, Future-Tech, Japan). Vickers Hardness $H_v$ was obtained by using different loads with holding time of 10 s.

$H_v$ was determined from $H_v = 1854.4F/L^2$, where $F$ (N) is the applied load, and $L$ (μm) is the arithmetic mean of the two diagonals of the Vickers indentation.

**TEM observation**. We used Argonne Chromatic Aberration-corrected TEM (ACAT, a FEI Titan 80-300ST TEM/STEM) with a field-emission gun to investigate the crystallographic orientation, high-resolution transmission electron microscopy (HRTEM) images and electron energy loss spectra (EELS) of the recovered samples. The ACAT is equipped with a CEOS spherical and chromatic aberration imaging corrector to allow a resolution better than 0.08 nm information limit. To prepare the TEM sample, we first retrieved the lonsdaleite from high pressure and temperature treatments and transferred the sample from the DAC chamber to a clean marble mortar with a tiny pin. Then, we crushed the recovered sample into powder by using the marble mortar and pestle. Finally, we dispersed the crushed powder onto a holey carbon grid. We also used focused-ion beam (FIB) technique to prepare plane-view and cross-sectional TEM specimens.

**X-ray and UV Raman measurements**. X-Ray Raman was performed at 16ID-D station at Advanced Photon Source at Argonne National Laboratory. A silicon (555) analyze crystal was aligned to receive the inelastic scattering intensity from the sample. The (555) Si crystal was set to reflect 10.175 keV x-rays; and the incident beam energy was tuned from 10.165 keV to 10.215 keV with energy step size 0.1 eV. UV Raman experiments on the quenched samples from HPHT experiments were conducted using a Renishaw inVia Raman microscope with a 325 nm HeCd laser at the Center for Nanoscale Materials at Argonne National Laboratory. This ultra-violet laser minimized the interference of fluorescence in Raman measurements. A laser power of 0.55 mW was used in the measurements in order to avoid any damages to the samples.

**Simulation**. To study the stability and vibration frequency of the proposed structure model, we carried out Density Functional Theory (DFT) calculations with plane wave basis sets as implemented in the VASP code[24,25]. All calculations were spin-polarized and carried out by using the gradient corrected exchange-correlation functional of Perdew, Burke and Ernzerhof (PBE) under the projector augmented wave (PAW) method with plane wave basis sets up to a kinetic energy cutoff of 500 eV. The

PAW method was used to represent the interaction between the core and valence electrons. The Kohn-Sham valence states (that is 2s 2p for C) are expanded in plane wave basis sets. For the geometry optimization calculations, the convergence criterion of the total energy was set to be within $1 \times 10^{-6}$ eV within the Gamma 8x8x8 K-point grid; and all the atoms and geometries were optimized until the residual forces became less than $1 \times 10^{-3}$ eV Å$^{-1}$. For the frequency calculations, the phonon values are obtained based on Density Functional Perturbation Theory (DFPT) to compute the Hessian matrix.

The MD simulations were performed using a semi-empirical long-range carbon bond-order potential (LCBOP) to model the interatomic interactions. The system simulated is a 8 x 8 x 8 supercell of the unit cell shown in Fig 3 and consist of 2048 atoms. The anisotropic pressure is applied using a Nosé Hoover barostat and the temperature is equilibrated using a Nosé Hoover thermostat. All equilibration simulations were performed for at least 2 ns.

AUTHOR INFORMATION
Corresponding Authors: yangwg@hpstar.ac.cn (W.G. Y.), jwen@anl.gov, (J.G. W.), maohk@hpstar.ac.cn (H.K. M.)

Notes
The authors declare no competing financial interest.

■ ACKNOWLEDGMENTS
Authors thank for David J. Gosztola at CNM for his help in UV Raman experiments. The authors like to thank the financial support from NSAF (U1930401) and National Nature Science Foundation of China (U1530402). HPCAT operations are supported by DOE, National Nuclear Security Administration under Award DE-NA0001974. 13BM-C operation is supported by COMPRES through the Partnership for Extreme Crystallography (PX2) project, under NSF Cooperative Agreement EAR 11-57758. This work, including the transmission electron microscopy, UV Raman, and MD simulation, was performed at the Center for Nanoscale Materials, a U.S. Department of Energy Office of Science User Facility, and supported by the U.S. Department of

Energy, Office of Science, under Contract No. DE-AC02-06CH11357. APS is supported by DOE−BES, under Contract DE-AC02-06CH11357 by UChicago Argonne, LLC. The gas loading was performed at GeoSoilEnviroCARS, APS, Argonne National Laboratory, supported by EAR-1128799 and DE-FG02-94ER14466.

Author Contributions: L.X. Y., Z.D. Z, D.Z. Z., Y.M. X. and W.G. Y. carried out the synchrotron experiment. J.G.W. and D.L. performed the TEM characterization. L.X. Y., W.G. Y., J.G. W. performed the experimental data analysis. K.C.L. performed the Raman spectrum calculation. H. T., B.M. Y, H.Y. G. performed the multianvil synthesis and Vickers Hardness measurements. Y.P. Y. prepared TEM sample by FIB. L.X. Y., K.C.L., J.G.W., S.S and S.S. carried out MD and DFT simulations. L.X. Y., J.G. W., W.G. Y., and H.K.M. wrote the manuscript. H.K.M. conceived and designed the project. All authors contributed to the discussion of the results and revision of the manuscript.

# Figure.1

(a) and (b) near c-axis XRD patterns on graphite sample taken at 13.2 and 20 GPa respectively. (c)XRD pattern of recovered sample after laser heated to 1400 K at 20 GPa. The diffraction spots in (a), (b)and (c)all show a clear hexagonal symmetry and are marked with different color cycles. (d)Transformation process from graphite to lonsdaleite. From left to right, alternating honeycomb layers of graphite slide and buckle to form lonsdaleite. Black solid and dashed lines connecting spheres of the same colors form new honeycombs of lonsdaleite, and red lines c-axis form bonds between lonsdaleite honeycombs. Blue and black spheres represent the close-packed planes of lonsdaleite in ABAB stacking sequence. (e)SAED pattern of the recovered specimen from a large area (1 mm in diameter) showing a hexagonal-like pattern, similar to XRD result in (a). (f)SAED pattern from a 200 nm area showing a 2-fold symmetry which can be well indexed by lonsdaleite [100] zone axis pattern of a single nanocrystal. (g)Schematic image of the epitaxial relationship between graphite and lonsdaleite.

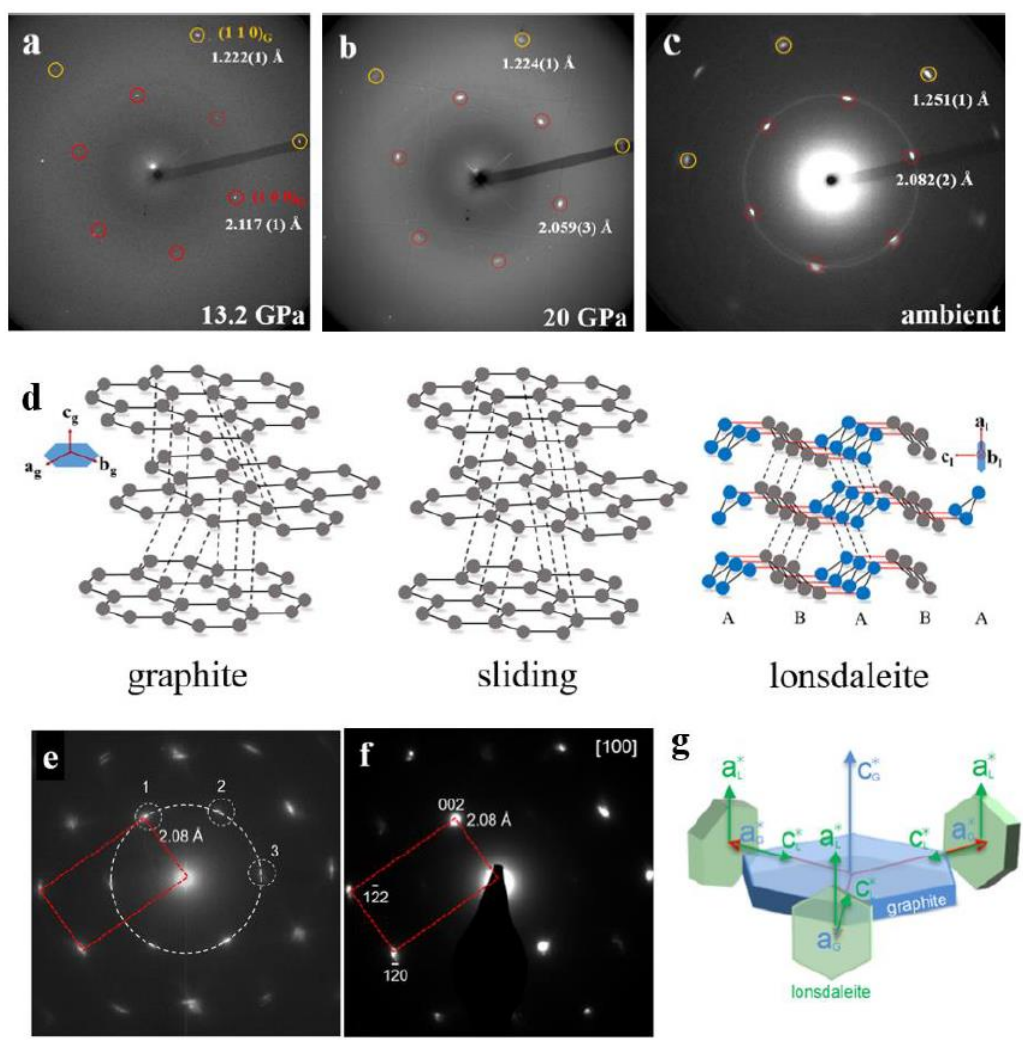

# Figure.2

(**a**) AC-HRTEM image of a lonsdaleite variant along [110] and (**b**) enlarged image showing two bond lengths of OA (~1.56Å) and OB (~1.47 Å) marked by orange and white circle, respectively. (**c**) Schematic diagram of the unit cell and tetrahedron in lonsdaleite with two bond lengths and angles.

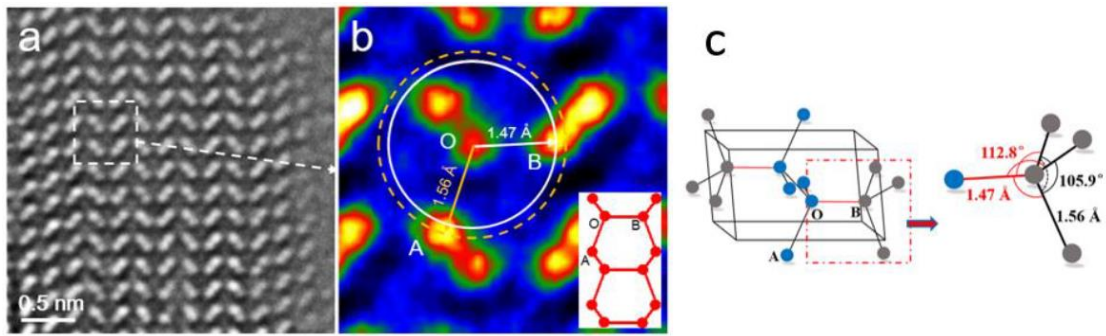

# Figure.3

(a) UV Raman spectrum of the recovered lonsdaleite sample showing three peaks. (b) Calculated Raman spectra of lonsdaleite based on bond length OB with fixed bond length OA (1.56 Å) show three Raman modes ($A_{1g}$, $E_{1g}$ and $E_{2g}$). Experimentally observed Raman vibration frequencies are indicated by red stars. (c) EELS spectra of carbon k-edge from lonsdaleite, diamond and graphite. (d) X-ray Raman spectra on recovered lonsdaleite samples with *ab* and *c* orientation. All results indicate 100% $sp_3$ bonding in the lonsdaleite.

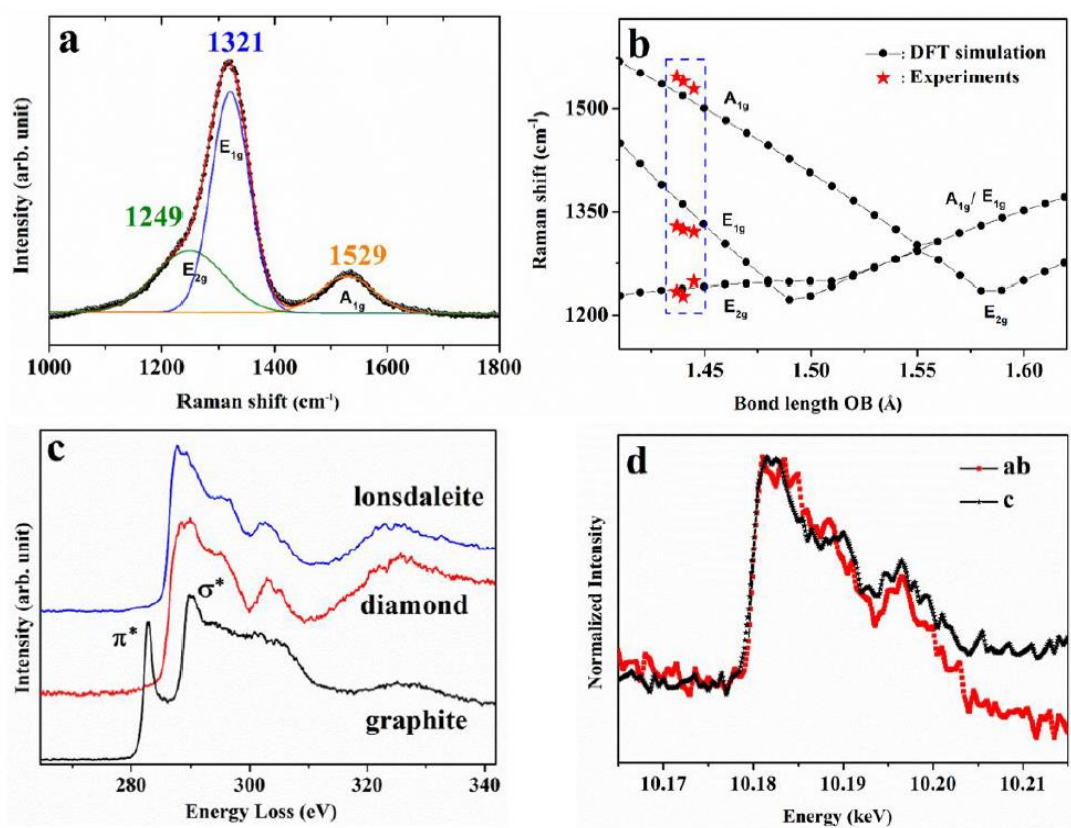

**Figure.4**

Vicker hardness tests of lonsdaleite along two directions with different load (N). Inset is a schematic image of the hardness test positions on sample and the C-C bond lengths in these two orientations also are labeled. The [1 0 0] direction of lonsdaleite is parallel with the c axis of the pristine graphite disk. Green symbols indicate indenting perpendicular to the lonsdaleite c-axis.

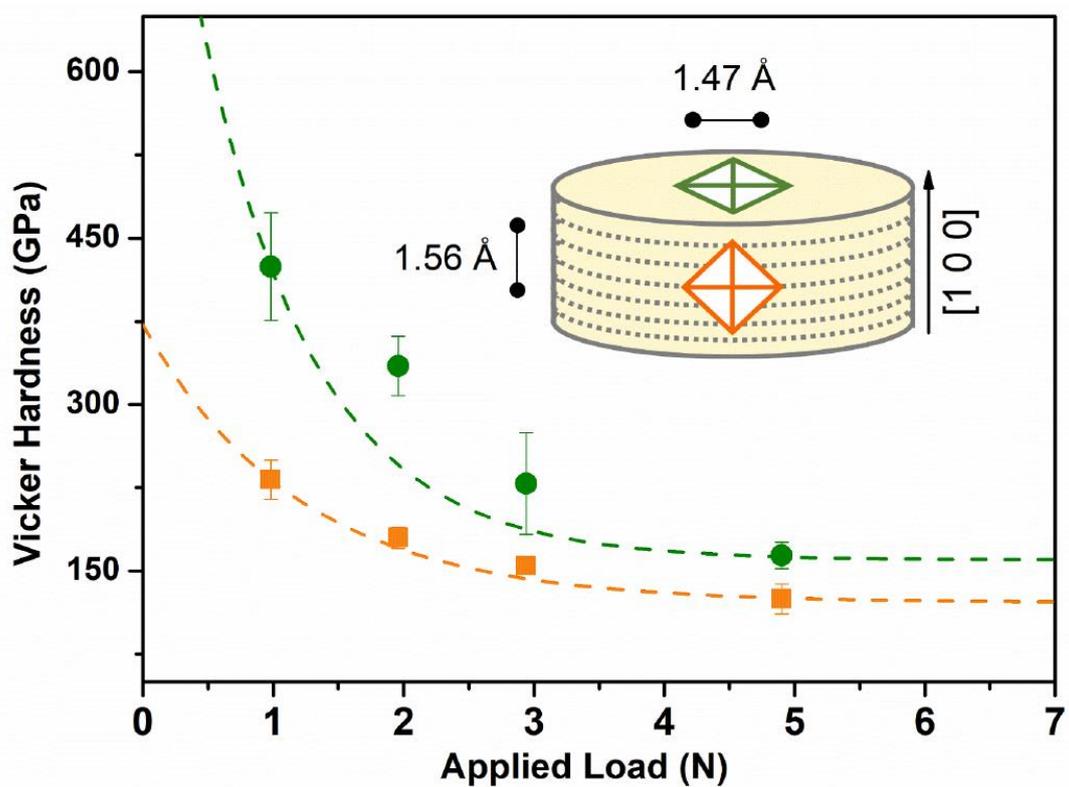

**Fig. SI-1** TEM images showing triple-twinned lonsdaleite. Dark-field TEM images show the recovered sample consists of three variants of domains. Fig SI-1B shows a six-fold like SAED pattern from the area in Fig. SI-1A. Dark field images (D-F) using diffraction spots 1, 2, and 3 show that there exist three variants. Each variant shows a 2-fold $[10\bar{1}0]$ diffraction pattern (Fig. SI-1C) with 120° apart.

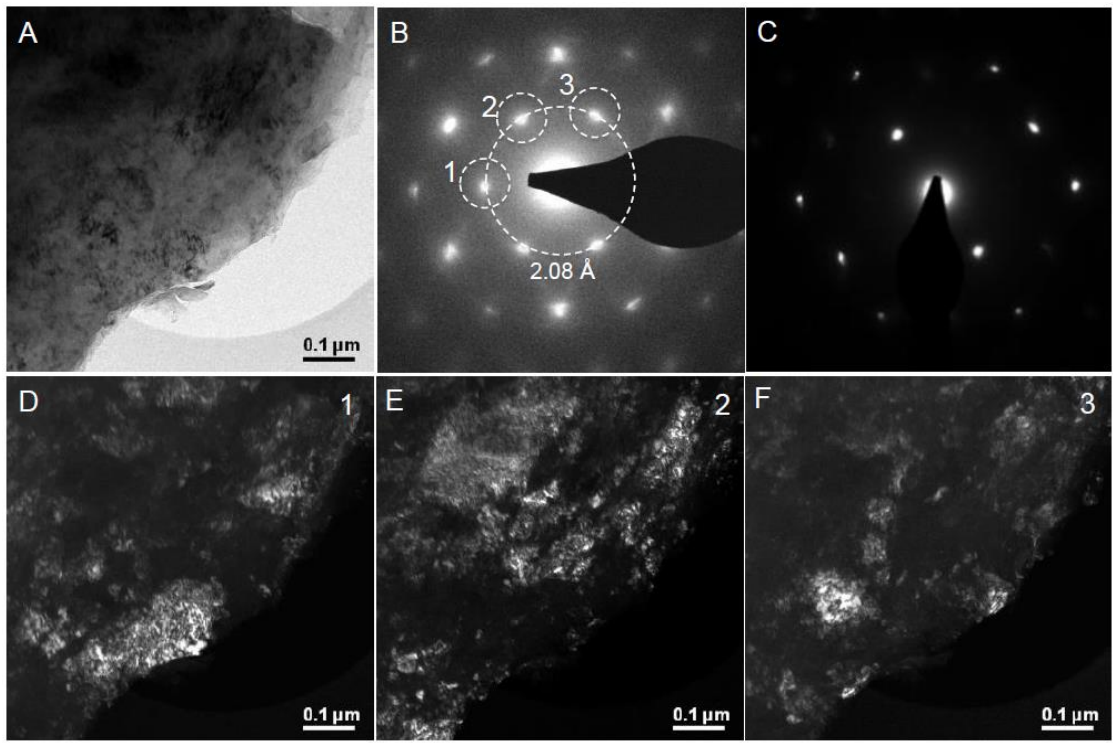

**Fig. SI-2(a)** Optical image of recovered lonsdaleite specimen in gasket chamber, which is completely transparent with some strip texture. **(b)** Pressure dependence of the volume of lonsdaleite. The solid blue line is fitted curve of the third-order Birch-Murnaghan (3BM) equation of state (EoS) with $V_0$=5.68(0.4) Å$^3$/atom, $B_0$=392(10) GPa, B'= 3.3(9). The red dash line is the previous result of the pressure dependent volume of cubic diamond.

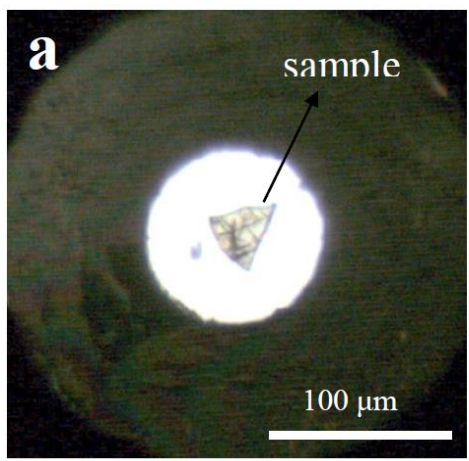
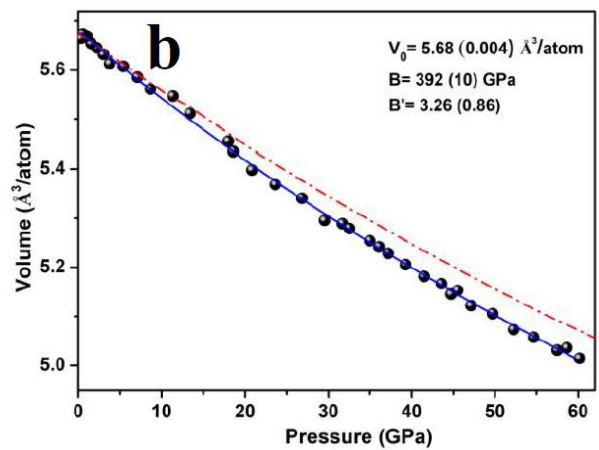

**TableSI-1.** d-spacing values of lonsdaleite from previous and our result with space group P6$_3$/mmc.

| Lonsdaleite | Boundy | Yagi | Our results (observed) | Calculated |
|---|---|---|---|---|
| 0 1 0 | 2.18 | 2.180 | 2.174 | 2.174 |
| 0 0 2 | 2.06 | 2.086 | 2.080 | 2.080 |
| 0 1 1 | 1.93 | | 1.969 | 1.927 |
| 0 1 2 | 1.50 | | 1.500 | 1.503 |
| 1 1 0 | 1.26 | 1.256 | 1.256 | 1.255 |
| 0 1 3 | 1.16 | | 1.170 | 1.169 |
| 0 2 0 | 1.09 | | 1.085 | 1.088 |
| 1 1 2 | 1.076 | 1.074 | 1.072 | 1.075 |
| P6$_3$/mmc | a = 2.52 Å  c = 4.12 Å | | a = 2.51 Å  c = 4.16 Å | |